\documentclass[aps,prb,twocolumn,showpacs,floatfix]{revtex4}

\usepackage{graphicx}

\begin{document}

\title{Magnetic field-induced transition in BaVS$_3$}
\author{P. Fazekas$^{1,2}$, N. Bari\v si\'c$^{3}$, I.
K\'ezsm\'arki$^{1}$, L. Demk\'o$^{1}$, H. Berger$^{3}$, L. Forr\'o
$^{3}$, and G. Mih\'aly$^{1}$}

\affiliation{$^{1}$Institute of Physics, \\Budapest University of
Technology and Economics,\\ H-1111 Budapest, Hungary\\
$^{2}$Research Institute for Solid State Physics and Optics\\
Hungarian Academy of Sciences,\\H-1525 Budapest, P.O.B. 49,
Hungary.\\
$^{3}$Institut de Physique de la Mati\'ere Complexe,\\EPFL,
CH-1015 Lausanne, Switzerland\\}
\date{\today}

\begin{abstract}
The metal-insulator transition (MIT) of BaVS$_3$ is suppressed
under pressure and above the critical pressure of $p_{\rm
cr}\approx 2{\rm GPa}$ the metallic phase is stabilized. We
present the results of detailed magnetoresistivity measurements
carried out at pressures near the critical value, in magnetic
fields up to $B=12$T. We found that slightly below the critical
pressure the structural tetramerization -- which drives the MIT --
is combined with the onset of magnetic correlations. If the
zero-field transition temperature is suppressed to a sufficiently
low value ($T_{\rm MI}\le$15K), the system can be driven into the
metallic state by application of magnetic field. The main effect
is not the reduction of $T_{\rm MI}$ with increasing $B$, but
rather the broadening of the transition due to the applied
magnetic field. We tentatively ascribe this phenomenon to the
influence on the magnetic structure coupled to the bond-order of
the tetramers.
\end{abstract}


\maketitle

The nature of the metal-insulator transition (MIT) of BaVS$_3$ has
presented riddles ever since its discovery a quarter of a century
ago.\cite{Massenet,neven} The ambient pressure MIT is spectacular in
the susceptibility cusp,\cite{Massenet,neven,graf,mihaly} but it is
also manifested in specific heat,\cite{shiga} thermal expansion,\cite{graf} resistivity\cite{Massenet,neven,graf,mihaly} and optical
reflectivity\cite{kezsmarki05, kezsmarki06} anomalies. The
measurement of transport properties under pressure\cite{neven,graf,forro,kezsmarki}
revealed that the insulating phase is
bounded by a line of MITs in the $p$--$T$ plane. The transition
temperature $T_{\rm MI}$ drops to zero at the critical pressure
$p_{\rm cr}\approx 2{\rm GPa}$ and the high-pressure $p>p_{\rm
cr}$ phase is metallic down to the lowest accessible temperatures.\cite{forro}

The order of the MI phase transition  was debated, and the options
of weakly first order, second order, and supercritical transition
have all been put forward. However, resistivity\cite{forro} and
in particular magnetoresistivity\cite{kezsmarki} measurements
under pressure led us to favor a second order
transition.  The character of the transition remains unchanged
under pressures up to $1.5$\,GPa as it always corresponds to a
sharp spike in the logarithmic derivative of the resistivity,
$d\log{\rho}/{\rm d}(1/T)$. This suggests a line of second order
MITs. One purpose of the present study to find out whether the
same phase boundary is preserved above $1.5$\,GPa extending up to
the critical pressure.

A critical line is possible only if it separates phases of
different symmetry, so the MIT in BaVS$_3$ must be at the same
time an ordering transition breaking some global
symmetry.\cite{pb02} However, the ambient-pressure MIT at 69K is
not accompanied by the appearance of detectable magnetic moments,
and the first two decades of search failed to reveal a
symmetry-changing distortion at $T_{\rm MI}$.\cite{footnote}
Finally, in 2002 Inami et al. reported the doubling of the unit
cell along the trigonal $c$-axis.\cite{inami} Since the unit cell
of the $T>T_{\rm MI}$ structure contains two V atoms along the
$c$-axis,\cite{structure}  the $T<T_{\rm MI}$ unit cell must
contain four, hence the symmetry breaking change corresponds to
tetramerization. More recent X-ray scattering measurements
reconfirmed the existence of four inequivalent V sites at low
temperatures.\cite{fagot2003,fagot2005,fagot2006} The X-ray
measurements were extended to 5K, well below the third and lowest
transition temperature $T_X=30$K of BaVS$_3$. The mysterious
$T_X$-transition leads to a low-$T$ phase with incommensurate
magnetic long range order,\cite{magn_str} and possibly orbital
order,\cite{naka97} but the symmetry of the lattice is held at
$T_X$.\cite{fagot2003,fagot2005,fagot2006}

The pressure range $1.5{\rm GPa}<p<p_{\rm cr}$, where $T_{\rm
MI}(p)$ drops below $\approx 20$K, had not been studied in detail.
It is expected that novel behavior may arise if the structural
transition and the magnetic transition eventually meet, or even
only get near. As $T_X$ was found to be pressure independent in
the pressure range investigated ($p<1~{\rm~GPa}$),\cite{private}
the $T_{\rm MI}(p)$ and $T_X(p)$ lines may merge, which could
change the nature of the MIT.

Another consideration is that even the well-studied low-pressure
MIT has a dual nature: it is the boundary of a phase with
spontaneous symmetry breaking (tetramerization), and at the same
time a metal--insulator phase boundary. Since the MIT coincides
with a symmetry breaking transition, the phase boundary may in
principle be continued to a zero-temperature quantum critical
point. On the other hand general arguments suggest that a
metal--insulator transition is likely to become first-order before
the predicted gap could become too small.\cite{mott} Furthermore,
it has been pointed out that the quantum critical behavior of an
MIT may be quite unlike the quantum criticality of a mere symmetry
breaking transition.\cite{imada05} BaVS$_3$ has both aspects,
thus it presents a situation of general interest but little firm
knowledge.

In this paper we investigate if there is a change in the nature of
the MIT at pressures close to the critical value. In order to get
detailed information about the range where $T_{\rm MI}(p)$ is
suppressed to zero, we carried out magnetoresistance measurements
at pressures where the characteristic energy of the applied field
is comparable to that of the transition temperature. Results
related to the MIT are presented here, while studies about the
high-pressure metallic phase will be reported in a companion paper.\cite{nevennfl}

Single crystals of BaVS$_3$ were grown by Tellurium flux method.
The crystals, obtained from the flux by sublimation, have typical
dimensions of $3 \times 0.5 \times 0.5 $ mm$^{3}$. The sample was
inserted into a self-clamping cell with kerosene as a pressure
medium. The pressure was  monitored in-situ by an InSb sensor. The
pressure was stable within 0.05 kbar in the pressure and
temperature ranges investigated in this work. The resistivity of
the single crystal was measured in a standard four probe
arrangement. The current was kept low enough to avoid the
self-heating of the sample. Magnetic fields up to 12 T were
applied perpendicular to the current, flowing along the
crystallographic $c$ direction.

\begin{figure}[t!]
\centerline{\includegraphics[width=0.7\columnwidth] {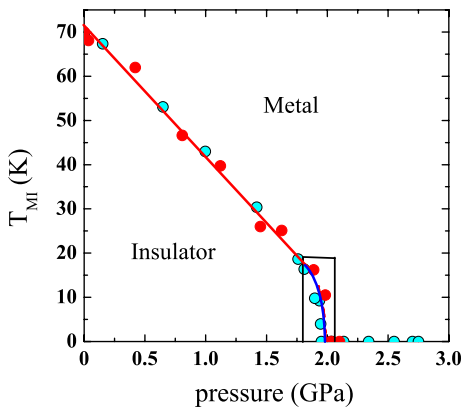}}
\caption{\footnotesize (Color online) Pressure dependence of the metal--insulator
phase boundary, as identified by the peak in the logarithmic
derivative of the resistivity. Results for two samples, marked by
different symbols, are presented. The boxed range indicates the
interval where results presented in the following figures were
obtained.} \label{fig:fig_1}
\end{figure}

The zero-field pressure dependence of the metal--insulator phase
boundary is shown in Fig.~\ref{fig:fig_1}. The boundary consists
of two distinguishable parts: $T_{\rm MI}$ decreases with $p$
almost perfectly linearly to about $p\approx 1.8$GPa where the
insulator becomes much softer, and the boundary begins to drop
steeply towards zero, signifying a new regime of the MIT. The
almost vertical drop of the $T_{\rm MI}(p)$ curve makes it
difficult to map this region in detail. The points in the boxed
range of Fig.~1 represent the data obtained by attempting to
fine-tune the pressure up to $p_{\rm cr}$. Even at $p=1.98$GPa
where $T_{\rm MI}$ is as low as to 7.5 K, the usual
characterization of the transition by $d\log{\rho}/{\rm d}(1/T)$
or magnetoresistivity curves yielded no striking difference from
the behavior seen up to 1.5GPa.\cite{kezsmarki} By these criteria,
the transition at $1.98$GPa is still second-order. However, the
considerable smearing out of the phase transition by magnetic
field  --~in addition to the downturn of the phase boundary~--
indicates that the character of the MIT changes in the parameter
range $p>1.8$\,GPa, $T\le 20$K (this domain is not sharply
delineated).

\begin{figure}[b!]
\centerline{\includegraphics[width=0.8\columnwidth]
 {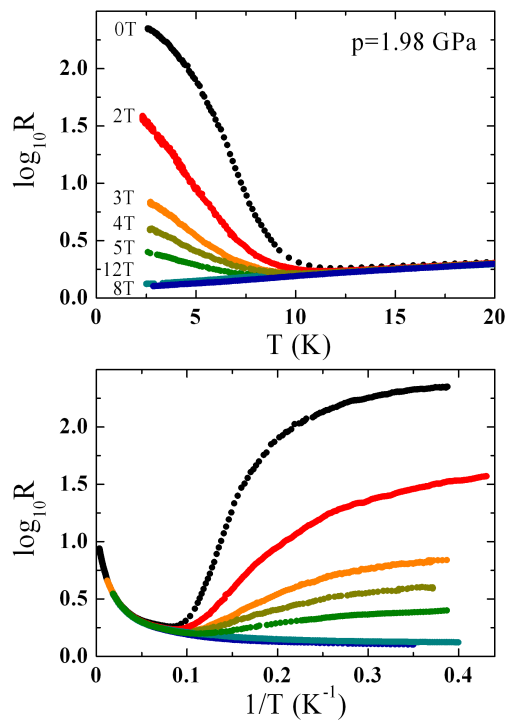}}\caption{\footnotesize (Color online) Temperature dependence of
 the resistance measured in presence of various magnetic fields at $p=1.98$GPa.
 At this pressure with increasing magnetic fields BaVS$_3$ undergoes a field-driven
 insulator-to-metal transition.}\label{fig:fig_2}
 \end{figure}

Former magnetotransport studies up to $1.5$\,GPa showed that the
low-field phase boundary follows a quadratic shape,\cite{kezsmarki} i.e.  $\Delta T_{\rm MI}(p,B)/T_{\rm
MI}(p)\propto [B/B_c (p)]^2$,
 where pressure dependent critical field and the zero-field
transition temperature scales together according to $\mu_{B}B_{c}
(p)=1.7 \cdot k_{B} T_{\rm MI}(p)$. In the case of the
$p=1.5$\,GPa measurement --~and for any lower pressures~-- no
broadening of the transition was observed up to the highest field
investigated $B_{\rm max}=12T$. Taking $T_{\rm MI}(p)$ as  the
characteristic energy scale, this field corresponds to $B_{\rm
max}\approx1/5T_{\rm MI}$. In contrast, in the high-pressure range
$p>1.8$\,GPa the transition is considerably smeared out by
magnetic fields smaller than $1/5\,T_{\rm MI}$, as shown in
Figs.~2. This implies that in the new regime of the transition,
i.e. in the boxed range of Fig.~\ref{fig:fig_1}, the insulating
phase became considerable less stable against the change of
pressure and the application of magnetic field. We guess that the
arising of an electronically soft state of matter is due to the
presence of competing orders.

\begin{figure}[b!]
\centerline{\includegraphics[width=0.87\columnwidth]
 {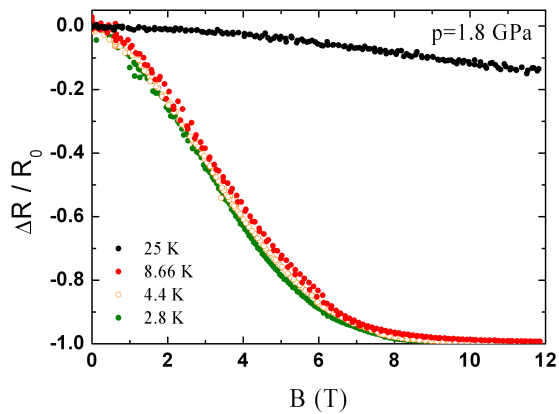}}\caption{\footnotesize (Color online) Magnetic field dependence of the
 resistance measured at various temperatures at $p=1.8$GPa. (The data are
 normalized to the zero field values.) The low temperature curves examples the
 field-driven insulator-to-metal transition. For comparison the
 magnetoresistance measured at $T=25$K is also shown.}\label{fig:fig_3}
 \end{figure}

We also found that if $T_{\rm MI}$ is suppressed below $\approx
15{\rm K}$ then the metallic state can be reached by applying
magnetic field. Such field-driven metal--insulator transitions are
shown in Figs.~2 and 3 for $p=1.98$GPa and $p=1.8$GPa, where at
zero magnetic field the transition temperatures are $T_{\rm
MI}=7.5$~K and $12$~K, respectively. Figure ~\ref{fig:fig_2}.
shows the temperature dependencies at various magnetic fields
(T-sweeps at $p=1.98$GPa). At low fields the low-temperature
resistivity is monotonically increasing with decreasing
temperature, indicating an insulating state. However, at high
fields the temperature dependence changes character and the ground
state of the system is metallic. Figure~\ref{fig:fig_3} present an
example for the magnetic field dependence of the resistance at
various temperatures (B-sweeps at $p=1.8$GPa). In both cases the
field-induced insulator-to-metal transition have taken place at
values well below $B_{c}(p)$ derived from the low-pressure scaling
rule. These would be $B_{c}=30$T and $18$T for $p=1.8$GPa and
$p=1.98$GPa, respectively. This also implies that the transition
in the boxed regime of Fig.~\ref{fig:fig_1} has a different
character, and we investigate this point in more detail below.

 \begin{figure}
\centerline{\includegraphics[width=0.6\columnwidth]
 {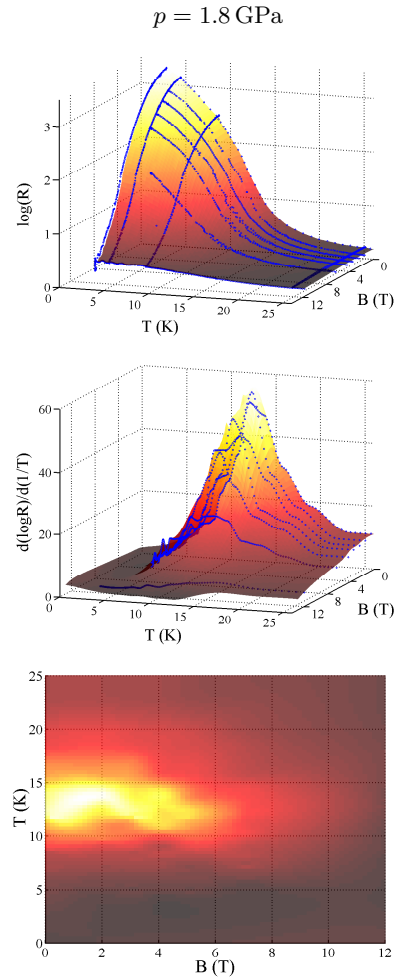}}\caption{\footnotesize (Color online) 3D plots obtained
 by temperature and magnetic field sweeps at $p=1.8$GPa (from light to dark: from highest to lowest values).
 {\sl Upper panel}: $\log{\rho}$ over the  B - T plane.
 The field gradually suppresses the insulating character. {\sl Middle panel}:
 Plotting $d\log{\rho}/{\rm d}(1/T)$ over the B - T plane shows
 that the field-induced transition gets weaker and less easily
 defined at higher fields.{\sl Middle panel}: Top view of the middle panel,
 i.e. the magnetic field dependence of the transition temperature derived from cusp-like anomaly in the temperature dependence of
$d\log{\rho}/{\rm d}(1/T)$}\label{fig:fig_4}
\end{figure}

The overall character of the field-induced transition is revealed
by the 3-dimensional plot of $\log{R}$ over the $B$--$T$ plane.
This is shown for $p=1.8$GPa in the upper panel of
Fig.~\ref{fig:fig_4}. Here the $R(B,T)$ surface is constructed
from the experimental $B$-sweep and $T$-sweeps; beside the results
shown in Fig.~\ref{fig:fig_3} this plot includes also temperature
dependencies at various magnetic fields. It is clear that at $B=0$
the system is most insulating  at low $T$, while the drive towards
the conducting state looks quite similar in $B$ and $T$
directions.

The transition temperature is generally identified with the position of the peak in the logarithmic derivative
of the resistivity. The middle and bottom panels of Fig.~\ref{fig:fig_4} show the $d\log{R}/d(1/T)$ curves over
the $B$--$T$ plane. The top view of the color-coded surface illustrates the magnetic field dependence of the
transition temperature. Though this might imply that a certain point $T_{\rm MI}(B)$ drops sharply as a function
of the magnetic field and a field-induced first order transition takes place\cite{spinpeierls}, the side view of
the surface reveals that that the main effect of the magnetic field is not the shifting of the cusp in the
derivative, but rather its general suppression.

\begin{figure}
\centerline{\includegraphics[width=0.6\columnwidth]
 {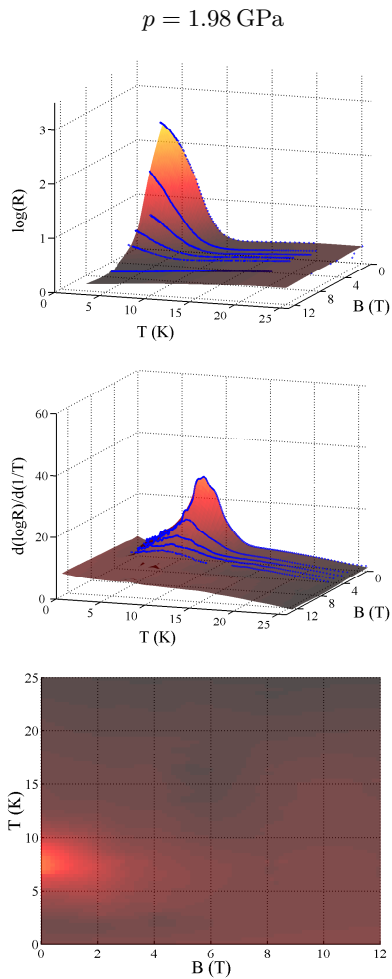}}\caption{\footnotesize (Color online) 3D plots obtained
 by temperature and magnetic field sweeps at $p=1.98$GPa.
 {\sl Upper panel}: $\log{\rho}$ over the  B - T plane,
  {\sl Lower panels}:
 $d\log{\rho}/{\rm d}(1/T)$ over the B - T plane (from light to dark:
 from highest to lowest values). A comparison to Fig.~4 reveals that with
 increasing pressure the insulating domain shrinks.}\label{fig:fig_5}
\end{figure}

The same analysis of the data recorded at $p=1.98$GPa
(Fig.~\ref{fig:fig_5}) shows that  with increasing pressure the
insulating domain shrinks, the transition temperature is shifted
down, but the general features remain unchanged. Again, by
increasing $B$, the transition becomes ill-defined indicating that
the character of the high-pressure MIT is definitely different
from that of the low-pressure (say $p<1.5$GPa) sharp transition.
We interpret the observed broadening as a sign that magnetic
ordering phenomena (related to those appearing at the low-pressure
ordering transition at $T_X$) begin to interfere with the MIT,
from $p\sim 1.8$GPa upwards or $T_{\rm MI}\sim 20$K downwards.

The suppression of non-ferromagnetic order by magnetic field is a
common phenomenon but it is not necessarily associated with the
broadening of the transition. There are many systems (including
ordinary antiferromagnets) which possess breakable symmetries even
in the presence of an external magnetic field. This residual
symmetry can be broken by an order parameter which is not induced
by the field. The critical temperature decreases with increasing
$B$ but the transition remains well-defined. An example is given
by the spin density wave system of $(TMTSF)_{2}PF_{6}$ which
continues to undergo a sharp phase transition around $T=$10K even
in magnetic fields as high as $B\approx 30$T.\cite{chaikin}

BaVS$_3$ is definitely not in the above category, the
metal-insulator transition is considerably broadened in fields as
demonstrated in Figs.~4 and 5. There are two possibilities: either
the order of the phase is induced by the field (thus in the
presence of the external field there is no symmetry breaking and
the transition becomes ill-defined), or the observed phenomena are
kinetic, and cannot be explained in the language of equilibrium
phase transitions.

Taking the first possibility, field induces in linear order the
corresponding components of spin and orbital momentum. This is
relevant if the low-$T$ phase BaVS$_3$ possesses a net
ferromagnetic moment (since the ferromagnetic moment may be very
weak, it may have escaped experimental detection). Even if the
zero-field ground state carries no total magnetization, field
could also induce higher-order uniform moments.  If one these is
mixed into the order parameter of BaVS$_3$ at $T<T_X$, the
observed broadening may once again follow.

Future experiments will decide whether the broadening of the
transition can be explained in terms of equilibrium phases, as
described above. However, we wish to point it out that there is an
alternative: namely that the broadening is a relaxation effect,
and thus not explicable  within the framework of equilibrium phase
transitions in ideal systems.

There is an independent observation showing that the character of
the MIT changes, and becomes mixed with, or preceeded by, magnetic
ordering phenomena in this $p$--$T$ regime; above $p\sim 1.8$GPa
hysteresis loops may appear both in $B$-cycles and in $T$-cycles
\cite{neven,nevennfl}. The shape of the hysteresis loops depends
on the sweep rate, suggesting that we are dealing with a
relaxation phenomenon. Hysteresis appears both above and below the
critical pressure, thus it is plausible to relate it to the
complicated long-period order which was observed by magnetic
neutron scattering \cite{magn_str,naka97,private}. Similar, so
called weak relaxation effects were observed in incommensurate
density wave systems and successfully interpreted in the picture
of a non-ideal system going through a hierarchy of pinned states
\cite{kriza}. Formally, the random impurity pinning leads to a
broad distribution of the relaxation times, and this results in
logarithmic, power law, or stretched exponential relaxation. The
finding that the ambient-pressure low-$T$ phase of BaVS$_3$
displays a magnetic order which is incommensurate both in the
basal plane and along the $c$-axis \cite{private} may provide the
key elements for an analogous relaxation mechanism. If an
incommensurate structure is locally pinned to defects, while the
optimal wave number is temperature dependent, the temperature
sweep yields an inhomogeneous state which try to relax towards the
optimal structure with a locally varying rate.

To summarize, we investigated the high-pressure low-temperature
region where the insulating phases are suppressed. The zero field
boundary was followed down to $T_{\rm MI}\approx 8$K. We did not
directly see a first order jump in either the temperature, or the
pressure, or the field dependence, but in the range $T_{\rm
MI}<20$K, $p>1.8$GPa  we observed a considerable smearing out of
the phase transition by magnetic field.  We suggest this is a sign
that the metal-insulator transition and magnetic ordering
phenomena appear combined in this ($p$, $T$) regime.

{\bf  Acknowledgments}. P.F. is greatly indebted to H. Nakamura
and T. Kobayashi for enlightening discussions and for their
communicating results prior to publication, to J.P. Pouget and S.
Bari\v si\'c for most valuable correspondence on the subject, and
to I. Kup\'ci\'c, K. Penc and K. Radn\'oczi for discussions on the
electronic structure of BaVS$_3$. This work was supported by the
Swiss National Foundation for Scientific Research and its research
pool "MaNEP" and by the Hungarian Scientific Research Fund OTKA
under grant Nos. TS049881, K62280, K62441 and Bolyai 00239/04.


\begin{references}
\bibitem{footnote} It was known that the MIT is associated with an anomaly
of the $c/a$ ratio.\cite{graf} This in itself would not change
the symmetry of the lattice.
\end{references}
\end{document}